# On the Spiral Structure of the Milky Way Galaxy

Yu.N. Efremov*

*Sternberg Astronomical Institute, Moscow State University, Moscow, Russia*

Abstract -- We consider the possible pattern of the overall spiral structure of the Galaxy, using data on the distribution of neutral (atomic), molecular, and ionized hydrogen, on the base of the hypothesis of the spiral structure being symmetric, i.e. the assumption that spiral arms are translated into each other for a rotation around the galactic center by 180° (a two-arm pattern) or by 90° (a four-arm pattern). We demonstrate that, for the inner region, the observations are best represented with a four-arm scheme of the spiral pattern, associated with all-Galaxy spiral density waves. The basic position is that of the Carina arm, reliably determined from distances to HII regions and from HI and $H_2$ radial velocities. This pattern is continued in the quadrants III and IV with weak outer HI arms; from their morphology, the Galaxy should be considered an asymmetric multi-arm spiral. The kneed shape of the outer arms that consist of straight segments can indicate that these arms are transient formations that appeared due to a gravitational instability in the gas disk. The distances between HI superclouds in the two arms that are the brightest in neutral hydrogen, the Carina arm and the Cygnus (Outer) arm, concentrate to two values, permitting to assume the presence of a regular magnetic field in these arms.

## 1. INTRODUCTION

The Milky Way star system being a spiral galaxy is beyond doubt since long ago, but the positions and number of the spiral arms are still a matter of debate. Spiral-arm indicators are well known: the arms concentrate the most massive gas clouds and at least young stars and clusters. Distances to the latter objects can currently be determined very precisely but only for those of them that are within 3—4 kpc from the Sun. During the most recent years, direct data have begun to accumulate on larger distances for sources of maser emission (related to young stars); they were obtained by means of very-long-base radio interferometry. Only a dozen of such distances are known so far, but even they can be helpful for improving distances to fragments of spiral arms whose existence and location were established from other data.

It is currently possible to make quite particular conclusions on spiral-arm positions at least for the outer Galaxy (i.e. for distances exceeding that of the Sun from the galactic center), based on data on clouds of neutral and ionized hydrogen, their velocities, and the Galaxy's rotation curve. For the inner Galaxy, it is often possible to choose between the short and large distances for clouds of atomic, molecular, and ionized hydrogen (both distances are derived from cloud radial velocities and the rotation curve) and also to get an impression concerning the spiral-arm locations.

*) efremovn@yandex.ru

It is known that three principal types of spiral galaxies can be distinguished. A spiral structure symmetric with respect of rotation around the galactic center, nown as grand design (GD) proper, is related to quasi-stationary spiral density waves due to deviations of the galaxy's gravitational potential from the axially symmetric one, as it is caused by the galaxy possessing a bar [1] or satellites; the best-studied examples of such a structure are the galaxies M51 and M81. The multi-arm, or kneed, structure (M101, NGC 1232) is probably transient and can be due to gravitation instability of galactic disks [2, 3]; it is sometimes considered a variety of the GD type, in particular because inner parts of such galaxies can feature regular spiral arms. Contrary to the first two types, flocculent (patchy) spiral galaxies (the best example is NGC 7793) contain many short fragments of arms that can be considered star complexes sheared by differential galactic rotation [4].

Data on other galaxies are known to demonstrate that HI superclouds, giant molecular clouds (GMCs) and young stars genetically related to them, and HII regions are located along spiral arms and often concentrated in giant star and gas complexes [5, 6]. The complexes in the spiral arms related to the all-Galaxy density wave probably are formed with magneto-gravitational instability due to regular magnetic field directed along the arm. Observational indications for this assumption were recently found in the north-western fragment of the S4 spiral arm in the Andromeda galaxy, just the arm with regular wave-shaped magnetic field along it detected long ago [see 7, 8].

We will try to achieve some conclusions on the all-Galaxy spiral pattern initially on the base of the hypothesis of its symmetric character, i.e. that the arms are translated into each other for a rotation around the galactic center by 180° (a two-arm pattern) or by 90° (a four-arm pattern). Such a symmetry is known in many galaxies with ``grand-design'' spiral patterns (for example, in M51). Evidence for the presence of this symmetry in our Galaxy was presented in many papers (for instance, cf. [9]).

Many years ago, de Vaucouleurs [10] classified the Milky Way as an SB(rs) galaxy. This classification is generally confirmed with the modern data on the distribution of atomic, ionized, and molecular hydrogen; we will demonstrate that these data lead to a four-arm spiral pattern within the solar circle. The two strongest gas arms are subdivided into HI superclouds, and we found them to be located at quasi-regular distances from each other along the arms, a possible indication of a regular magnetic field in these arms.

The data on the location of rich young clusters and the regions of active star formation agree with this concept. The arms traced by HI assume a kneed character in the outer part of the Galaxy, but in general, they continue the flow of the inner arms as they are represented with the four-arm scheme of the Galaxy.

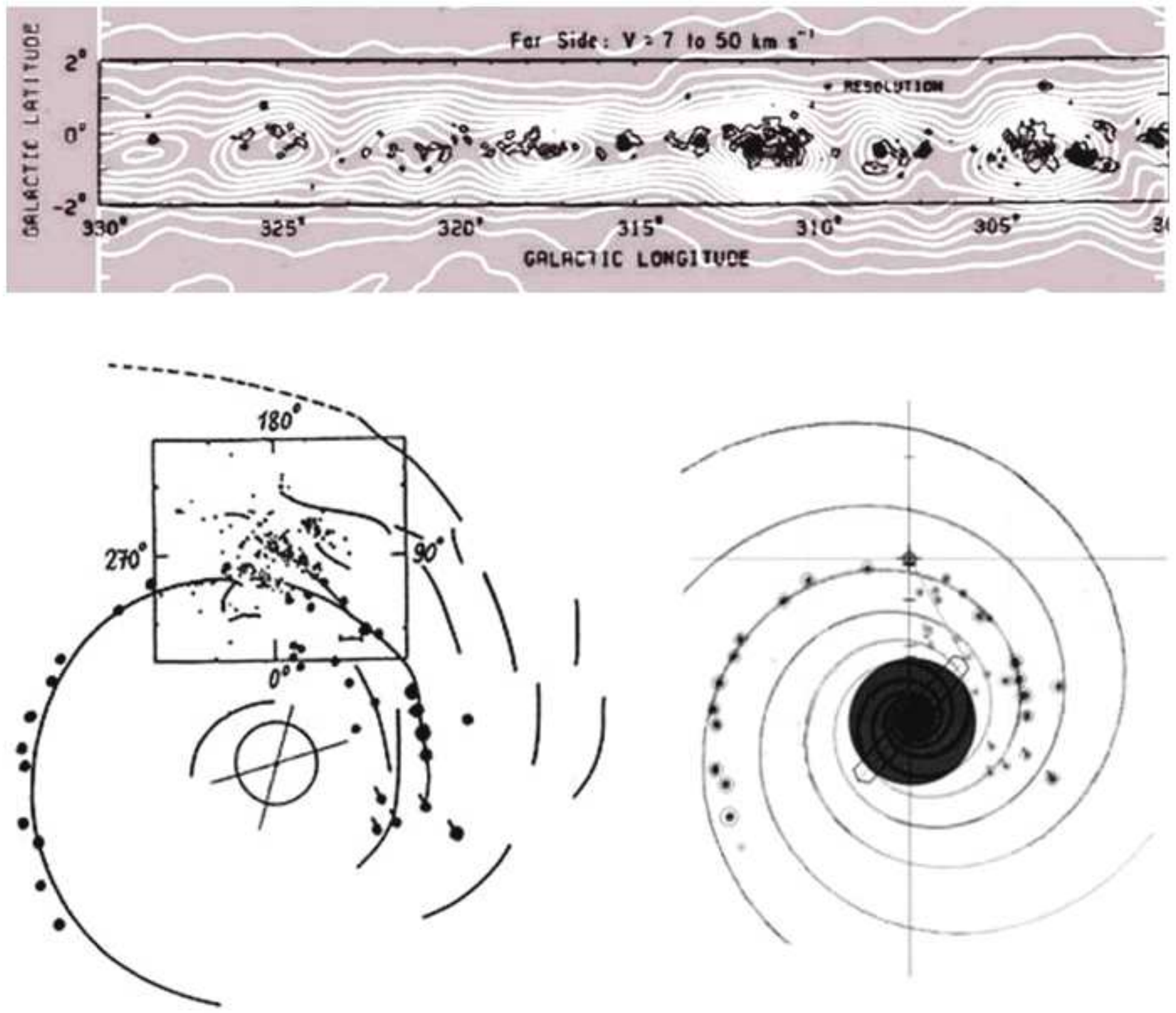

Figure 1

Top: the HI superclouds (light isodensity curves) and molecular clouds (thin black curves and dots) in a part of the Carina arm.
Bottom: the scheme of the HI arms from Weaver [16] and HI superclouds (large dots) in the Carina arm and in the II quadrant (left); the Carina--Sagittarius arm and its image repeated 3 times after rotation by 90° (right). See text.

## 2. THE CARINA--SAGITTARIUS ARM

A concentration of young clusters and OB~associations in the solar neighborhood was detected long ago in three fragments of spiral arms, known as the Carina--Sagittarius, Perseus, and Cygnus--Orion arms (on the history of this problem, see, for example, the book [5]). The first two of these fragments are, beyond doubt, parts of the all-Galaxy spiral pattern, while the Cygnus--Orion arm appears to be a bridge between the first two arms, probably beginning as a spur originating at the Sagittarius arm.

The key issue is the position of the Carina--Sagittarius arm, reliably determined from data on the distribution and velocities of the giant clouds of molecular hydrogen (GMCs). They indicate the presence of a single Sagittarius--Carina (Sgr--Car) arm, its length being about 40~kpc and its pitch angle, about $10^{\circ}$ [11]. As a rule, these clouds are located in the cores of supergiant clouds, or superclouds (SCs), of atomic hydrogen; this was noted in [11] and is especially evident when overlaying Figs. 7 and 15 from [12], as it is shown in Fig. 1 (top). Quite striking is the regularity of distances between the superclouds; it was first noted by Efremov [13] and later confirmed [7, 8, 14].

A large part of the Carina arm is outside the solar circle, and thus the kinematical distances to the gas clouds tracing the arm are determined uniquely. These distances are also free of considerable random errors. The distances to closer HII regions are determined from their exciting hot stars rather than kinematically, from radial velocities and the rotation curve. This regions do not contradict the general smooth shape of the arm and thus can be used to ``bring to anchor'' the position of the arm as a whole. Besides, the line of sight in the 280° - 300° longitude range is directed along the arm, so that uncertainties of distances to objects in this longitude range do not change the position of the arm.

Based on the data by Grabelsky et al. [11, 12] for the IV quadrant (they correspond to a limited range of velocities, from 7 to 50 km/s, and hence of distances) and on the data by Elmegreen and Elmegreen [15] for the whole I quadrant of galactic longitudes, we derived [13, 14] the distribution of superclouds in the Car--Sgr arm displayed in Fig. 1 (bottom left). For the IV quadrant, we read the longitudes of supercloud centers from Fig. 15 in [12]; the distances to them were assumed the same as those to the largest GMCs (given in [11]), located within the HI superclouds in the sky plane (Fig. 1, top). The data for the I quadrant were taken directly from [15].

## 3. THE FOUR-ARM SCHEME OF THE GALAXY

It appears from Fig. 1 (bottom left) that the position of the Carina--Sagittarius arm defined by HI superclouds is in a good agreement with the Galaxy's spiral pattern plotted by Weaver [16] on the base of HI data.

The spiral pattern in ``grand-design'' galaxies is strictly symmetrical with respect of rotation around the galaxy center by 180° (this was demonstrated especially evidently for M51 [17]), and if a second pair of arms is present, these arms are located between the two arms of the first pair. Thus, it is possible to restore the whole spiral pattern knowing the position of a single arm, like a paleontologist is able to restore the whole image of a fossil animal from a single bone. Admittedly, significant deviations of the pattern from the regular one, as well as gaps and additional arms are observed in most cases, but the presence of a bar in our Galaxy (cf. [18]) means that the dominant pattern should be a more or less regular two-arm or four-arm structure.

If the assumption that our Galaxy belongs to the GD type is correct, the Carina--Sagittarius arm should overlap the other arms if rotated around the galactic center. Figure 1 (bottom right) shows four logarithmic spiral arms with a pitch angle of 12°. The positions of all the arms in this spiral pattern are uniquely determined with the location of the Carina--Sagittarius arm, rotated three times by 90° around the galactic center. Note that most studies based on the distribution of HI clouds and HII regions in the whole Galaxy result in a four-arm spiral structure.

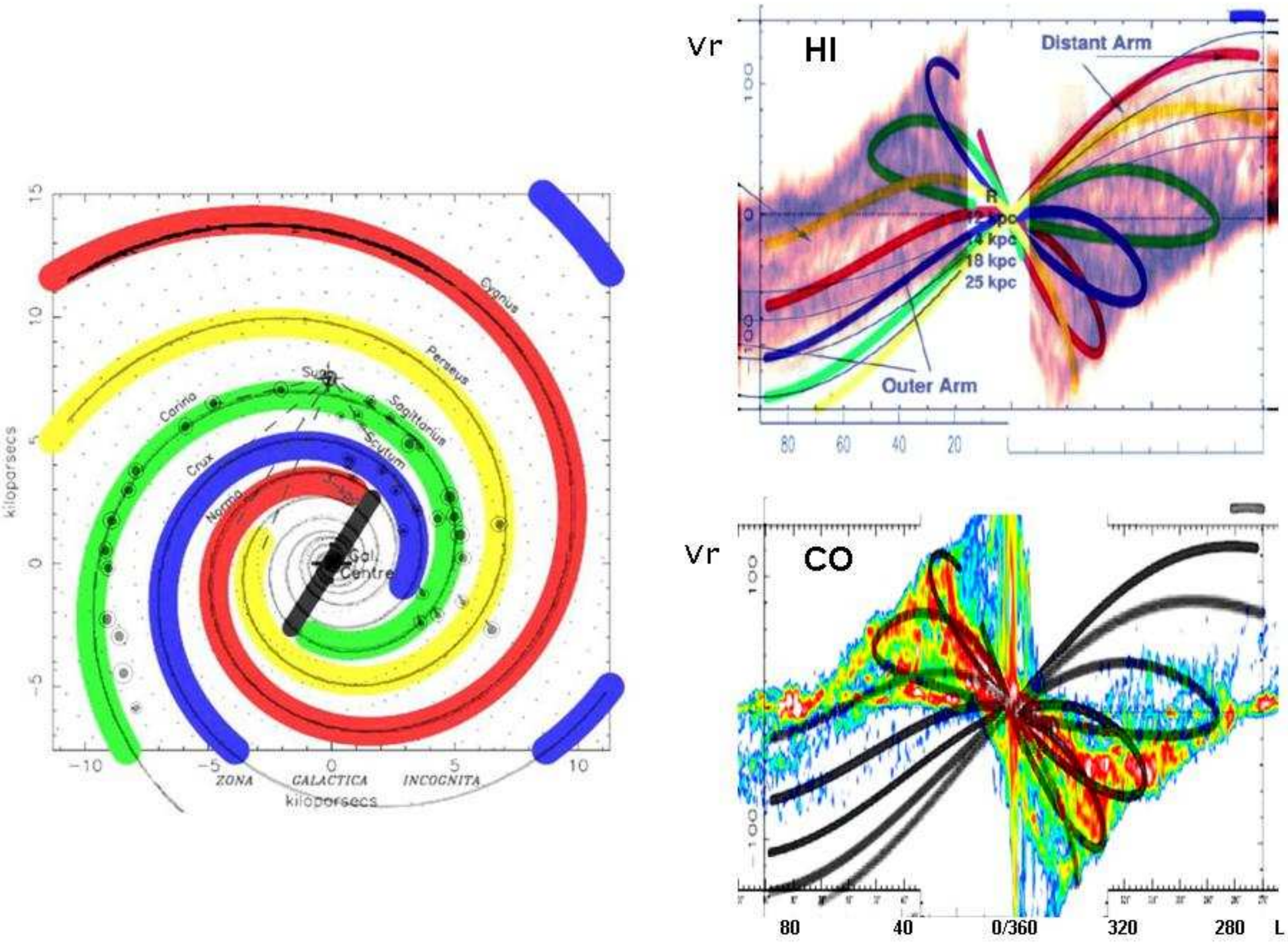

Figure 2

Left: the four-arm scheme of the spiral structure, based on [14, 9, 8].
Right: observed HI (top) and CO (bottom) radial velocities (the observational data are from [19, 20]), with overlaid computational results by Vallée [9].

Figure 2 (left) demonstrates that this scheme of the spiral pattern, based solely on the shape of the Carina--Sagittarius arm and on the assumption of the Galaxy's spiral pattern being symmetric, perfectly fits that derived by Vallée [9] who summarized numerous studies devoted to this problem.

The right side of Fig. 2 displays the relations of the radial velocity on longitude for each of the arms in the I and IV longitude quadrants derived by Vallée [9] based on his scheme and the assumption of the constant velocity of the galactic rotation (220 km/s). We overlaid his curves with the observed radial velocity curves for neutral hydrogen (Fig.2, top right; the data from [19]) and for CO (Fig. 2, bottom right; the data from[20]). We may consider the agreement good; the data for the Carina arm agree especially well. Note that Fig. 2 demonstrates that the Outer arm found in the IV quadrant in [21] should be considered a continuation of the Cygnus arm.

The positions of the tangents to the arms shown in the Vallée's scheme (Fig. 2, left) generally agree with the kinematic-arm turn over longitudes, apparent in the observation data for the HI and CO velocities (Fig. 2, right). It should be noted that, according to Fig. 7 in Luna et al. [22], arm tangent-point longitudes in the IV quadrant are located just within the distances to the galactic center where the rotation is closer to that of a rigid body (i.e. corresponds to a constant angular velocity). According to Elmegreen [23], such a reduction of the amount of shear (i.e. the rotation being rather solid-body and not differential) should be observed within the spiral density waves. Similar fragments of rigid-body rotation were also detected in the rotation curve based on northern data.

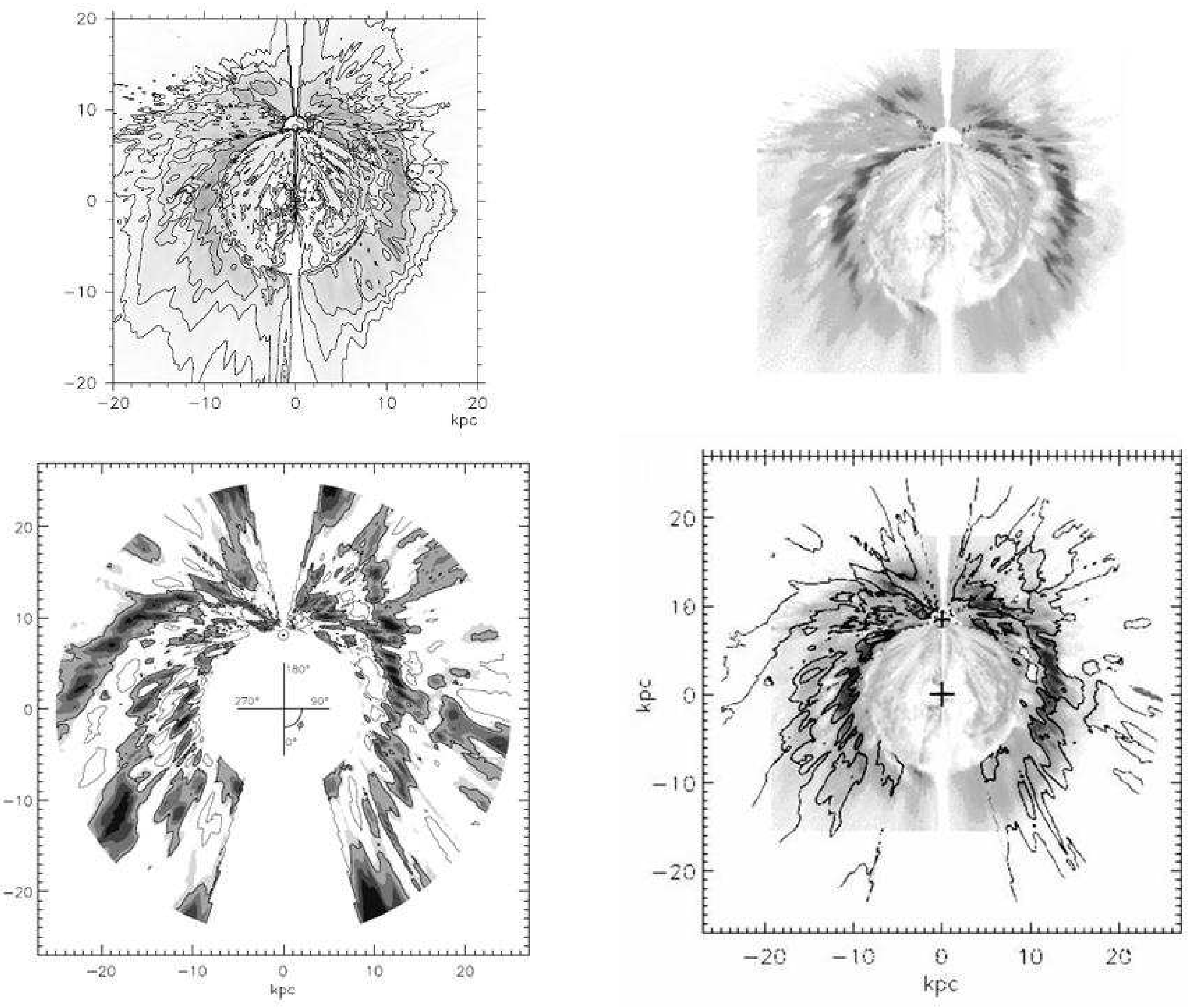

Figure 3

Top: the distribution of HI in the Galaxy according to [24] (left) and the raster image of the same map extracted from the pdf file using the Adobe Photoshop image editor (right).
Bottom left: the distribution of HI in the outer galactic regions according to [26].
Bottom right: the top right map, overlaid with boundaries of enhanced hydrogen density areas (as compared to the mean local density), determined using data from [25, 26].

Figure 3 displays the maps of the distribution of HI in the Galaxy based on virtually the same observations and the same rotation curve (Nakanishi and Sofue [24] -- top left; Levine et al. [25] -- bottom left). Both studies reveal almost similar, very well-expressed Carina and Cygnus arms, but their results differ strongly for the outer parts of the Galaxy. These outer regions of the spiral structure in the II and IV quadrants, with a low density and strongly deviating from the galactic plane, are merely absent in Nakanishi and Sofue [24].

Levine et al. [25, 26] were able to trace them using unsharp masking treatment, which emphasizes density contrasts at its intermediate and low levels.

Figure 3 (bottom right) demonstrates that the agreement between the two studies is in fact excellent. Closer to the galactic center, the positions of the Carina and Cygnus arms in the two papers coincide, the shape of the Cygnus arm and of the regions of enhanced HI density (superclouds) in this arm being virtually identical. The explanation is apparently that the two studies used approximately the same galactic-rotation velocities (220 km/s) for distance determinations (in the 7—15 kpc range from the center). Note that the Carina arm and the Cygnus arm are equally strong within the I and IV quadrants.

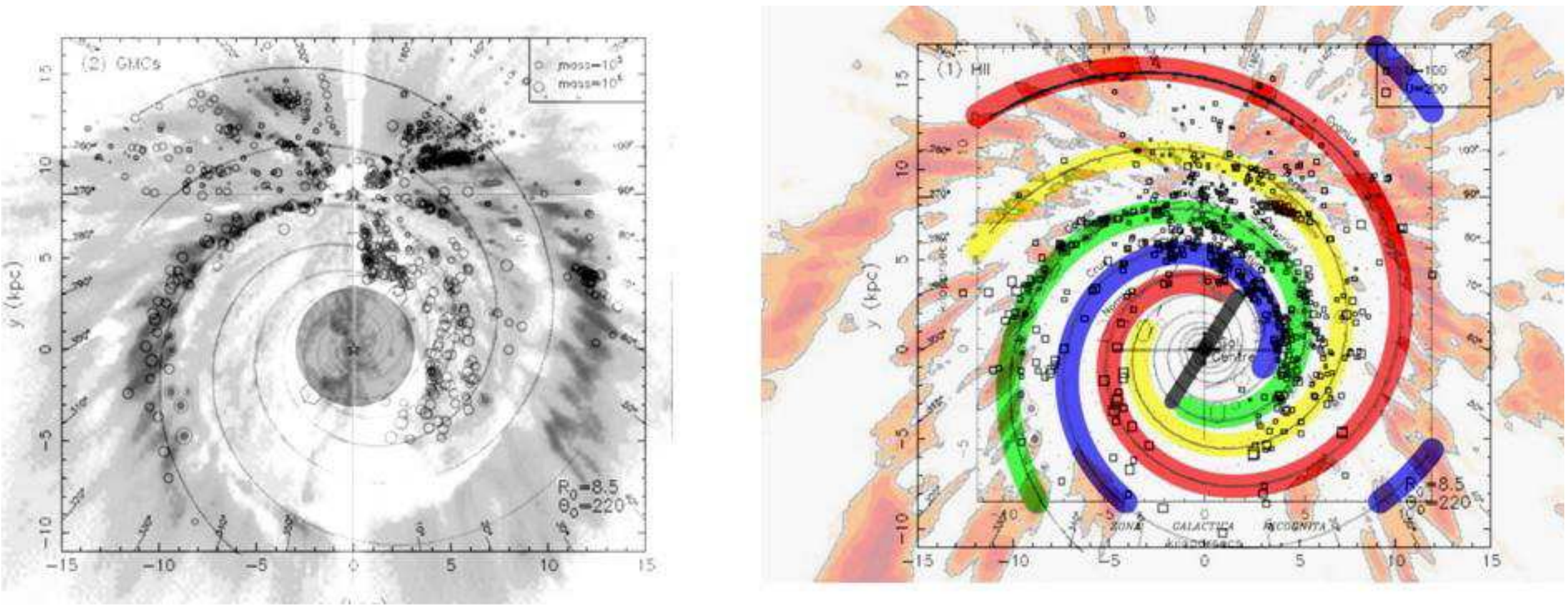

Figure 4

Left: our four-arm scheme overlaying the GMC positions (circles) according to [27] and the HI distribution according to [24].
Right: the scheme of the outer HI arms according to [26], overlaying the positions of HII regions according to [27], and Vallée's four-arm scheme.

We see from Fig. 4 that, overlaying the four-arm scheme from Vallée [9] or Efremov [14, 8] with the HI distributions derived by Nakanishi and Sofue [24] (left) and by Levine et al. [25, 26] (right), we get an excellent agreement for the Carina arm and a satisfactory agreement for the Cygnus arm (Fig. 4). As already mentioned, this scheme gives positions of the arms identical to those from three rotations of the Carina--Sagittarius arm by 90° around the galactic center. The four-arm Vallée's – Efremov's scheme is in a good agreement with the data on distributions of the GMCs (Fig. 4, left) and HII regions (Fig. 4, right), taken from the currently most complete data collection in Hou et al. [27].

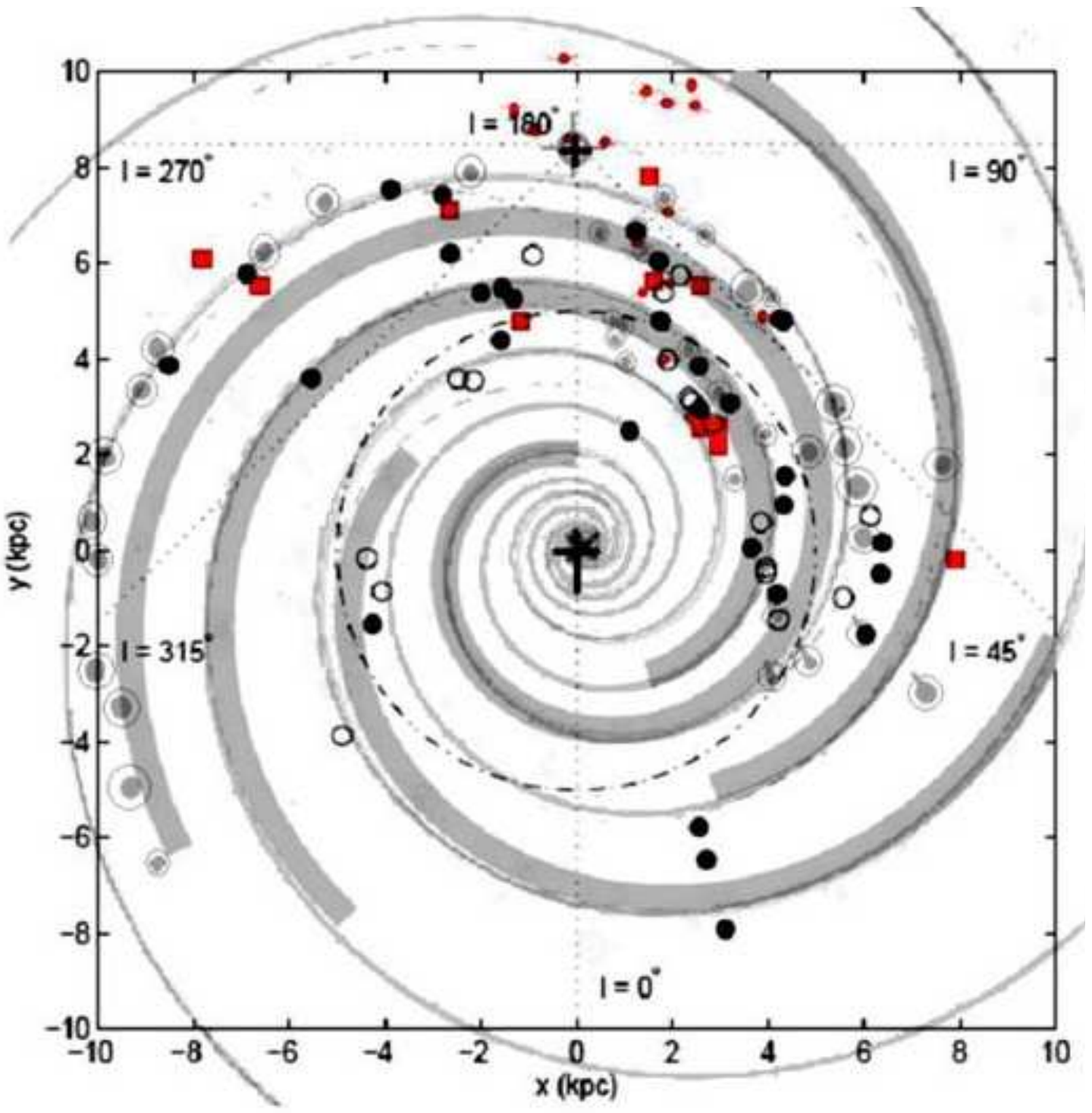

Figure 5

Our four-arm scheme (dotted circles are HI superclouds) overlaid with the distributions of rich young clusters from [28] (red squares); of star-formation regions with maser sources (red dots), their distances derived from VLBA data [29]; of the largest star-formation regions (open circles), their distances derived from IR data [30]; and with the scheme of the spiral structure suggested in [31] from CO data (thick spiral arms).

Figure 5 shows that the data on the distribution of young rich clusters and of giant star-formation regions in the Galaxy excellently agree with the four-arm scheme. This scheme also successfully represents the positions of a dozen of young rich clusters recently discovered in the Galaxy from IR observations and plotted according to data from [28]; the scheme also agrees with the distances to maser sources (related to young high-mass stars) determined from their annual parallaxes using very-long-base radio interferometry [29]. Also, this scheme better agrees with the positions of giant star formation regions (presented in [30] on the base of Spitzer GLIMPSE Survey data), than the model (also a four-arm one) suggested by the authors of [31] using their data on the distribution of molecular hydrogen. Note however that the scheme in [31] differs from our scheme only with respect of the position of the parts of the Carina--Sagittarius arm, most distant from the center (Fig. 5).

## 4. FOUR ARMS OR TWO ARMS?

Consider now data on Cepheids and open star clusters, objects with rather reliable distances, concentrated in spiral arms, as it follows from their distribution in other galaxies. Unfortunately, their data can be treated as complete only for distances within about 3~kpc. It appears from Fig. 6 that Cepheids (ages of most of them being between 50 and 100 million years) concentrate to the Carina arm and, to a lesser extent, to the Perseus arm;

they are also present in the bridge connecting the Sagittarius and Perseus arms (earlier often called the Cygnus--Orion arm). Younger clusters and O~associations reliably trace these three fragments of arms, known for a long time. The data on Cepheids based on extensive series of observations by L.N.Berdnikov were taken from [32]; the data on young clusters, from [33]; and the data on OB~associations, from [34]. Figure 6 (bottom left), taken from [35], presents the distribution of open clusters compared to fragments of spiral arms with the pitch angle of 6°.

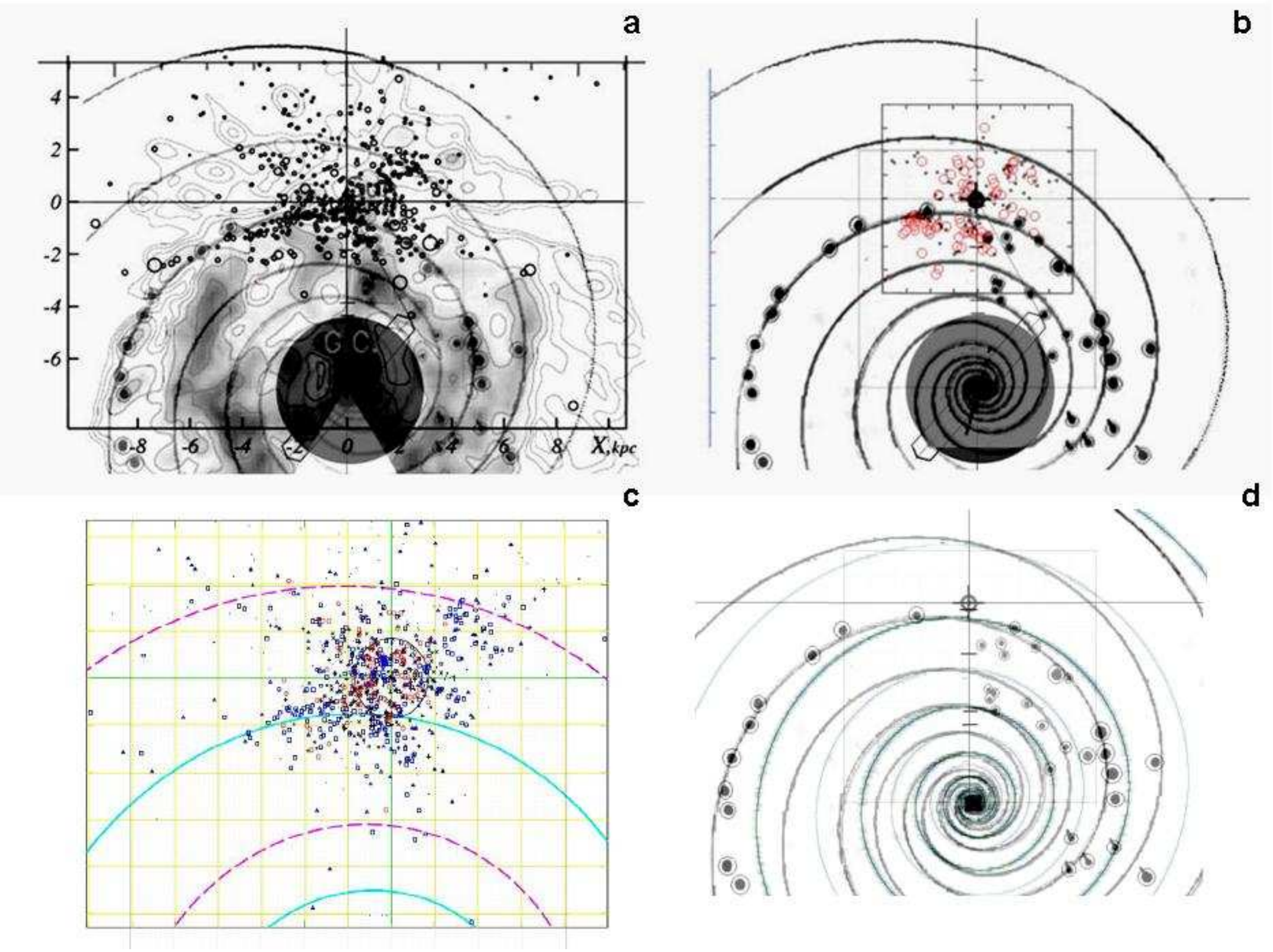

Figure 6

Top: the four-arm scheme overlaid with the distributions of CO according to [31] and of Cepheids according to [32] (left) and with the distributions of young clusters according to [33] and of O-associations according to [34] (right). The probable position of the Galaxy's bar is marked.
Bottom: the distribution of open clusters overlaid with the two-arm spiral structure with the pitch angle of 6° according to [35] (left) and comparison of the arm positions in four-arm and two-arm schemes (right). See text.

In general, Fig. 6 demonstrates that the clusters and Cepheids (observable only in the solar vicinity) do not permit a definite choice between the four-arm scheme (with the pitch angle of 12°) and the two-armed scheme (pitch-angle 6°). To compare both options, the position of the spiral pattern is uniquely fixed by the condition that the Sun is at the distance of about 0.7--1.0 pc from the axis of the Carina--Sagittarius arm. It seems however that the distribution of gas, especially in the outer parts of the Galaxy, evidences for the four-arm scheme.

Figure 7 displays the galactic distributions of HI (top) and CO (i.e. molecular hydrogen; bottom) with overlaid spiral patterns, the two-arm one (left) and the four-arm one (right).
We see that the pattern in the left half of Fig. 7 strongly contradicts the direction of the HI spiral arms in the outer parts of the Galaxy, while the four-arm pattern in the right half of this Fig. agrees with all the regions of enhanced density, those of HI as well as of CO, satisfactorily. Only in four-arm scheme the inner arms are connected with the most outer HI arms, found recently in {25, 26] (see also Fig. 4).

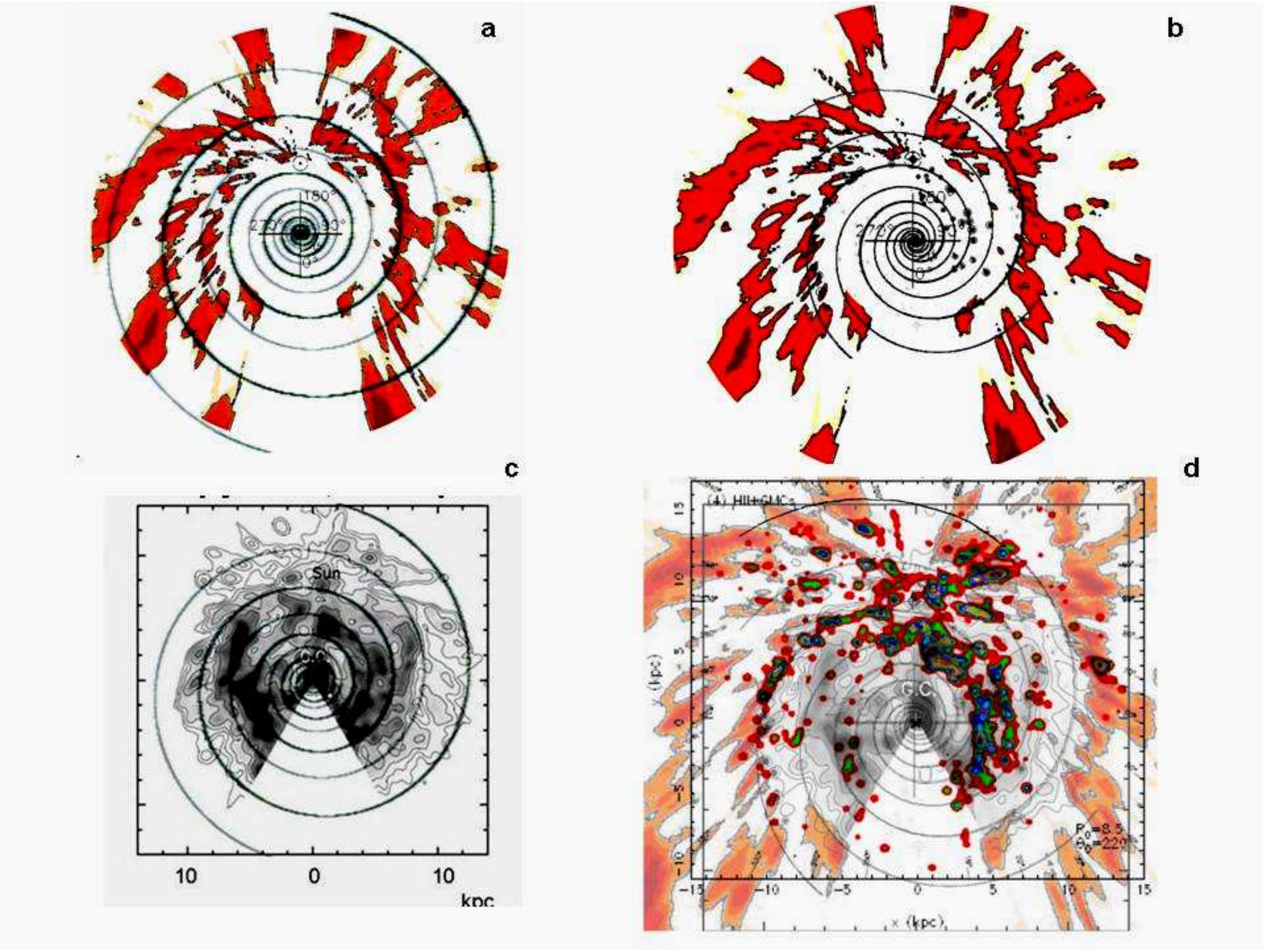

Figure 7

Left: the two-arm spiral structure overlaid with the distributions of neutral hydrogen according to [26] (top) and of molecular hydrogen according to [31] (bottom).
Right: the four-arm scheme overlaid with the distribution of HI from [26] (top) and the same map appended with the CO distribution according to [31] and with averaged positions of HII regions and GMCs according to [27] (bottom).

Thus, the scheme of four almost symmetric arms with the pitch angle about 10° --12° meets no contradictions at the distances from the galactic center up to 9—10 kpc. Note that the pitch angle about 6° that follows from the two-arm scheme is close to the lowest observed for other galaxies (4°), while 12° is close to the value following from the correlation found by Kennicutt [36] between the development level of the bulge (i.e. the Hubble spiral-galaxy subtype) and the pitch angle. According to this correlation, the pitch angle of 6° needed for the two-arm model, is possible only for Sa galaxies, while there is no doubt that the Milky Way galaxy belongs to a later type (presumably Sbc). Note however that ``grand-design" galaxies display two arms more often than four arms.

According to four-arm scheme, the Cygnus arm (Outer arm), like the Carina--Sagittarius arm, originate near the tips of the bar, whose most probable direction and size are indicated in Fig. 6 (and 1) according to [18]. The Cygnus arm is equally strongly expressed in neutral hydrogen as the Carina arm, but it is much weaker in molecular hydrogen (CO), corresponding to the decrease of the fraction of molecular hydrogen with respect of neutral hydrogen with the distance from the center, observed for all galaxies. Thus, the star formation intensity is much lower in the Cygnus arm. It appears from Fig. 4 that only few star-formation regions are contained within this arm, concentrating to the highest-density HI superclouds. According to Levine et al. [25], the Carina arm is somewhat below the mean plane of the Galaxy, while the Z--coordinate of the HI layer in the II quadrant, beyond the Cygnus arm, is already 2 kpc at the distances about 15 kpc, continuing to increase with distance from the center.

## 5. SPIRAL ARMS IN THE IR RANGE

Let us now consider the results of the study of longitude distribution of old red stars detected in the medium IR--range by the Spitzer Space Telescope [37, 38]. The authors of the cited papers detected an excess (by 20% --30%) of red stars in the direction of tangents to the Scutum--Centaurus arm (in the terms used by Vallée and other authors, the Crux--Scutum arm) and assumed that these two directions corresponded to an arm originating at the end of the Galaxy's bar closest to us. No excess of this kind was detected along the tangents to the Sagittarius--Carina arm, and thus this arm possesses (as discussed above) an enhanced density of gas but not of old stars (Fig. 8). Note however that young red supergiants are bright in the IR range too, and though their spatial density is much lower than for old red giants, their relative contribution to red-star counts is expected to increase at large distances. This problem should be studied. James and Seigar [39] present data indicating that the contribution from red supergiants in the K--band is from 3% to 33%.

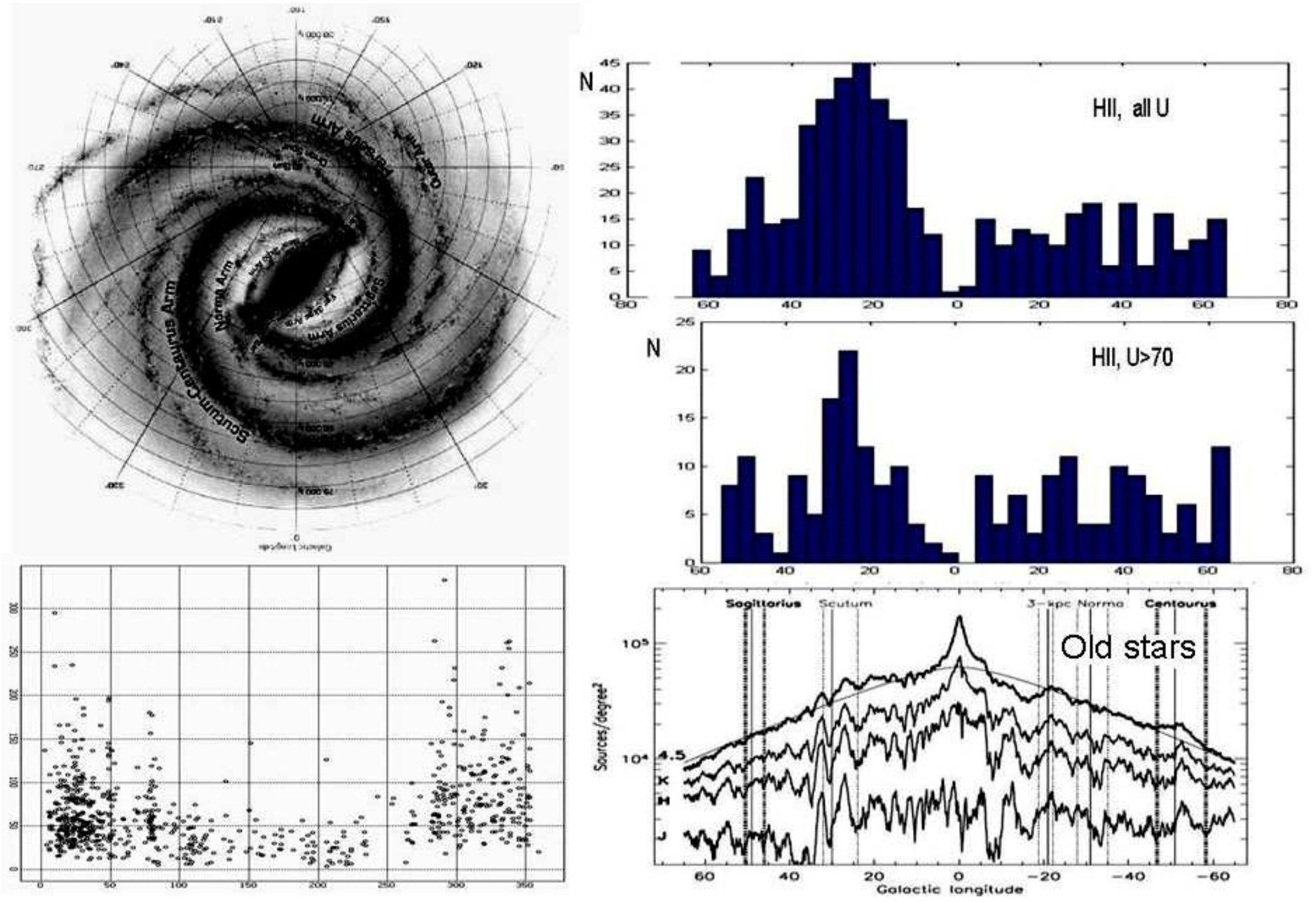

Figure 8

Left: the scheme of the spiral structure presented in [38] (top)
and the longitude distribution of HII regions based on the data from[27] (bottom).
Right: the longitude distribution of HII regions (top) compared to
the distribution of red stars (bottom) taken from [38].

We have already noted that the Carina--Sagitarius arm, as well as the Outer (Cygnus) arm, are better expressed in HI than all other arms. The data on HII regions collected in Hou et al. [27] were used to plot Fig. 8 (top right) that demonstrates a maximum in the HII-region distribution observed along the tangent to the Scutum arm, appearing sharper for stronger HII regions (Fig. 8, bottom left). This indicates that the arm in question (Scutum--Crux = Scutum--Centaurus) concentrates gas as well as stars, young and old.

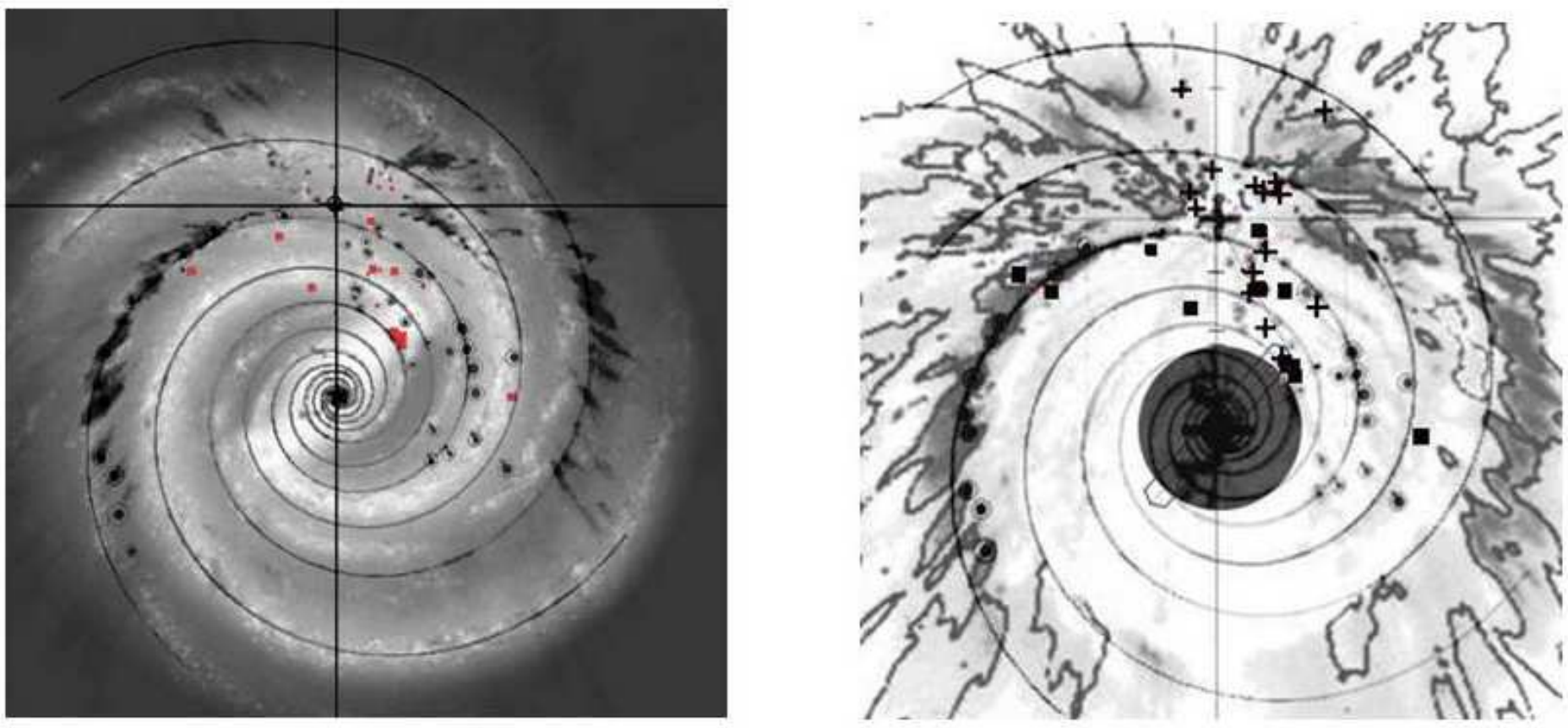

Figure 9

Left: the four-arm scheme overlaid with the scheme (white areas) from [38], similarly aligned, and with the distributions of HI (black areas) according to [24] and of young high-mass clusters from [28] and star formation regions from [29], both in red.
Right: the scaled bottom right map from Fig. 3. The same clusters (squares) and SFRs (crosses), as in left part, are added. Compare with Fig. 5.

At the first glance, the scheme of the Galaxy's spiral arms (top left of Fig. 8) taken from Churchwell et al. [38] contradicts the Vallee--Efremov scheme, but Fig. 9 demonstrates that it is not the case. The positions of the spiral arms virtually coincide in both schemes; there is a disagreement only in the assumptions concerning particular arms originating at the tips of the bar.

Note that Fig. 9 shows especially clearly that young rich clusters are located near the spiral arms and that a group of four such clusters is situated near the tip of the bar. It is also known for other galaxies that star formation regions tend to positions near the tips of the bar.

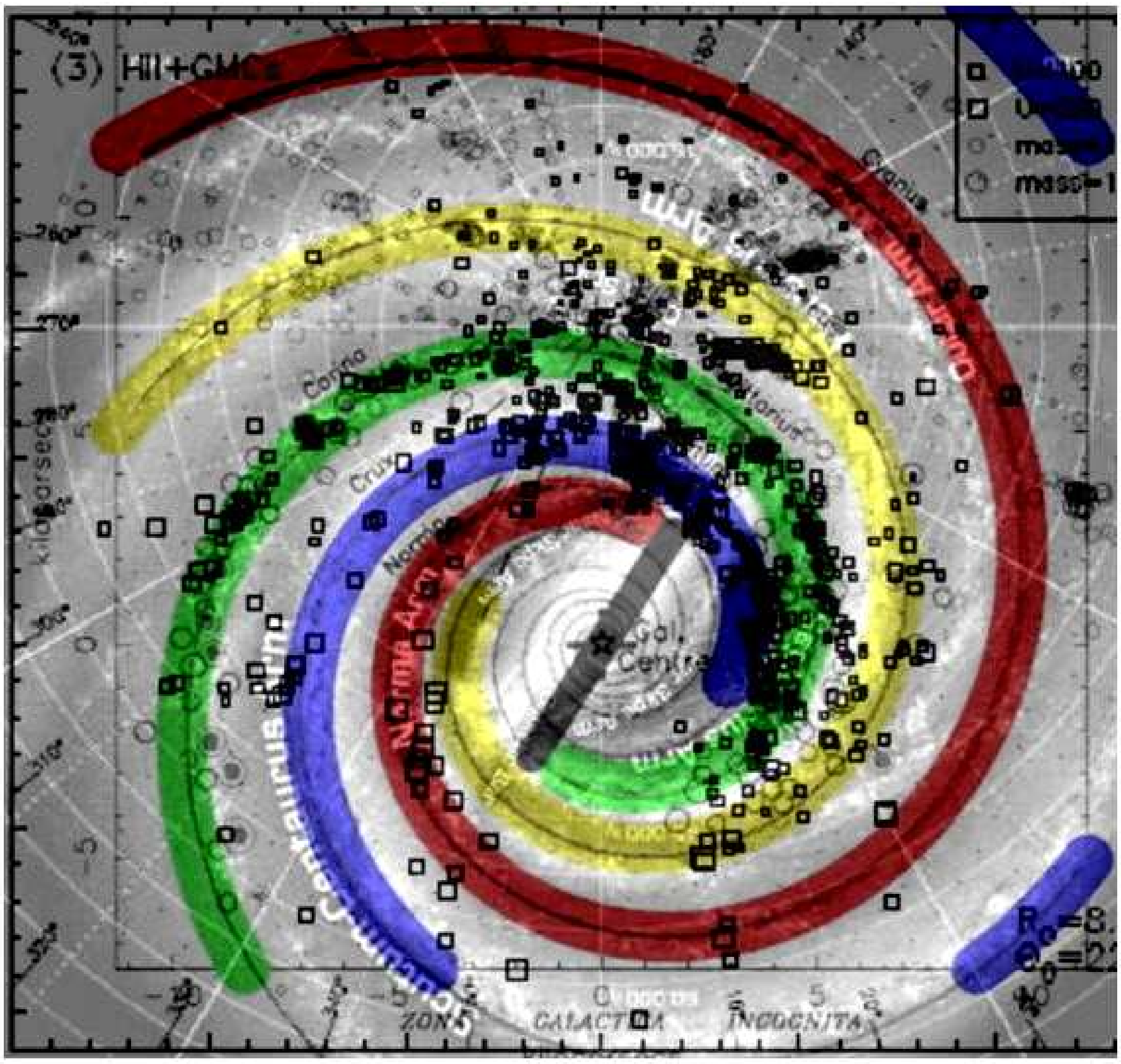

Figure 10

The four-arm Vallée--Efremov scheme, with overlaid positions of HII regions (squares) and GMCs (circles) according to [27], and with the spiral-structure scheme (white areas) according to [38].

Figure10 is a combined presentation of the spiral structure, as it is traced with the four-arm scheme suggested by us and by Vallée, with molecular clouds and HII regions according to [27], and with star counts in the IR—range [38]. Once again, it shows a good agreement of the existing structure schemes of the inner Galaxy. The presence of young stars in the Carina--Sagittarius arm is beyond doubt, as its follows from the concentration of molecular hydrogen as well as of HII regions in it, but the Cygnus (Outer) arm, equally bright in HI, contains only a single region concentrating clouds of both molecular and ionized hydrogen.

It should be specially noted that, according to the four-arm scheme, these two strongest arms of the Galaxy originate near the opposite ends of the bar, as it is usually the case, and they both consist of a chain of superclouds. The superclouds are extended along the line of sight because of the large velocity dispersion within them (and hence of the distances determined from the rotation curve); an uncertainty in their mean distances is also possible, but the superclouds' galactic longitudes (and hence the distances between them along the arm) are determined with certainty.

It would be interesting to check what modifications of the galactic rotation curves at distances from the center larger than 8 kpc will be needed for the northern and southern sky, in order to achieve an exact coincidence of the Cygnus and Carina arms after the rotation by 180°.

## 6. HI SUPERCLOUDS IN THE CARINA AND CYGNUS ARMS

Both arms best expressed in HI reveal a clear fragmentation into superclouds that were first detected by McGee and Milton [40]. Their maps of HI emission in the outer Galaxy exhibit a chain of superclouds, their mean mass being $10^7$ suns and their longitudes corresponding to the superclouds seen in Figs. 1 and 3 in the Carina and Cygnus arms. Later on it was found [7, 13, 14] that the distances between superclouds concentrated to two preferred values, most probably because of their formation under the influence of a magneto-gravitational instability developing along the arm. A confirmation of this hypothesis is the recent discovery [8] of similar distances between star complexes along just that particular fragment of one of M31 spiral arms, where Beck et al. [41] detected a regular wave-like magnetic field along the arm.

We display in Fig. 11 the distribution of distances between the superclouds measured along the arm, in the units of the Sun's distance to the center, in Fig. 11 (blue - for the Carina arm; red - for the Cygnus arm). Adding the Cygnus arm makes it possible to confirm the tendency we earlier found [13, 14] in the data for the Carina--Sagittarius arm: the superclouds tend to positions along the arm separated predominately by distances D of 0.1Ro or 0.2Ro between them, Ro being the galactocentric distance of the Sun.

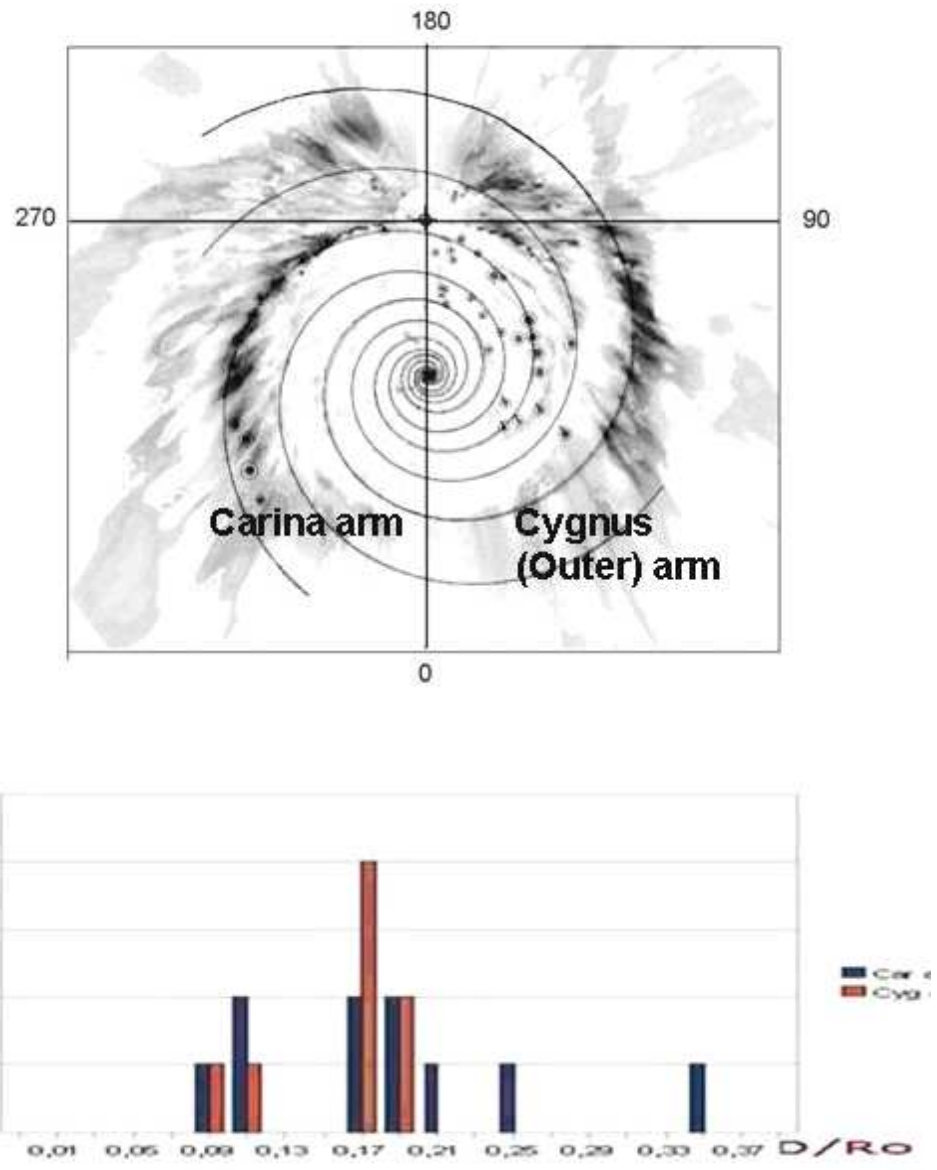

Figure 11

Top: the four-arm scheme [7] with overlaid distribution of HI according to combined data from [24, 26].

Bottom: the distribution of mutual distances between HI superclouds in the Carina and Cygnus (Outer) arms according to [7, 8].

Figure 11 shows two cases of approximately twice larger than average distances between superclouds in the Carina arm, but in both these cases we find a GMC between these superclouds (GMCs No. 21 and No. 38 in Table 2 of Grabelsky et al. [11]). Elmegreen and Elmegreen [42] describe a similar situation for four of the 22 galaxies where they found quasi-regular chains of star complexes (called HII regions in the cited paper). The most probable explanation is that the complexes are formed *only* near extrema of the regular wave-like magnetic field along the arm, though there are no complexes at some of the extrema. However, as mentioned above, star formation regions less extended than complexes are found at the corresponding positions also in such cases. This is a rather instructive, though indirect, indication that it is just the regular magnetic field along the arm and the magnetic and gravitational instability related to it that are responsible for formation of the quasi-regular chains of gas and star complexes.

The earlier detected correlation between the positions of star complexes and magnetic-field parameters in a segment of a spiral arm in M31 [7, 8] confirms the correctness of our assumption, but it should be noted that the presence of a magnetic field, regular along arms, remains not confirmed for our Galaxy. The quasi-regular distribution of star complexes along arms was detected by Elmegreen and Elmegreem [42] for 22 spiral galaxies, while a total of some 200 galaxy images from the Palomar atlas were selected for their search for star-complex chains (B.Elmegreen, private communication). We can conclude that regularity of this kind is rare. In 7 cases, chains of complexes could be found in only one of the arms [42].

It can be assumed that cases of regular magnetic field along an arm are rare and are limited to some fragments of arms, thus explaining why chains of star and gas complexes are rarely met in galactic arms. However, it is certainly possible that, in the presence of such a field, some other conditions are needed as well for regular chains of complexes and/or their paternal superclouds to be formed [see discussion in [8]).

## 7. DISCUSSION

In the previous sections, we mainly considered the distribution of gas and young stars in the Galaxy. Let us now address the data permitting to judge on the morphology of its spiral pattern taken in all its integrity.

The doubtless presence of long arms (in particular, the Carina--Sagittarius arm), identified by gas as well as by young (and probably old stars too), indicates that our Galaxy does not belong to the type of flocculent spirals, but a choice between its being a wave (``grand-design") galaxy or a multi-arm galaxy is difficult (and maybe not needed, as we will see below).

In a perfect case, in GD galaxies where spiral density waves are in operation, a gradient of ages across an arm (far from corotation; cf. [43]) is expected. Strict symmetry with respect of rotation by 180° is often observed, as it was clearly demonstrated for M51 in [17] (i.e. the galaxy center is an axis of symmetry of the second order). There is no symmetry in multi-arm galaxies, at least far from their centers, and there is no reason to expect a gradient of abundances or ages across their arms because the potential well is in the middle of the stellar arm in such galaxies [44].

An important argument for the assumption that inner regions of the Galaxy's spiral structure belong to those generated with the all-Galaxy density wave is provided by the rigid-body-rotation fragments observed in the rotation curve of the Galaxy just at the distances from the center corresponding to the location of spiral arms [22]. Earlier Elmegreen [23] demonstrated from the wave theory of the arms that, within the arms, rotation should indeed be closer to the rigid-body character than to the differential one. Luna et al. [22] found three fragments of rigid-body rotation within the Galaxy's spiral arms in the IV quadrant and confirmed the conclusion by Elmegreen that such a character of the rotation favored the formation process of star-generating gas clouds.

Note that the inner arms in M101 and NGC 1232 are more symmetric, they reveal dust lanes along the inner edges of the stellar arms; a kneed multi-arm structure appears only far from the center. The arms identified by Levine et al. [25, 26] in the outer Galaxy from the excess of HI density over the local level (Figs. 3 and 7) obviously consist of straight fragments, and they about exactly continue, in the III and IV quadrants, the inner four-arm spiral structure.

It is possible to notice from the figures in [44, 45] that kneed arms can be formed due to an internal gravitational instability (unrelated to any influence from a bar or from close neighbors) developing in a galaxy's gas disk (see also [2] and [46]) and resulting in a transient multi-arm spiral pattern. Its individual arms appear and disappear, but the general spiral pattern is seen at any time. Such arms, in variance to those due to a spiral density wave, which rotating as a rigid body, should not reveal an age gradient across the arms. However, in our Galaxy indications of such a gradient are known in the solar neighborhood [47] and there exist kinematical indications that the parts of the Carina—Sagittarius arms closest to us agree with the theory of spiral arms as density waves [see, for example, 48].

According to Chernin [49], the kneed spiral structure can originate in a spiral shock waves whose front tends to become flat, so that the arm becomes subdivided into straight rows, as he noticed for M51. In such a case, there obviously should exist an age gradient across the arm (from dust and molecular hydrogen to HII regions and then to older and older stars), and it is really observed just in the straight fragment of the S4 arm in M31 [8, 50].

We saw that the inner spiral arms had a pitch angle about 12°; however, this angle is evidently larger for the outer parts (the spirals are not so tight). The ``global" value for the pitch angle of the Galaxy's spiral arms can be estimated using the relations with the mass of the central black hole and with the velocity dispersion in galactic bulges that correlates with these parameters [51].

Figure 12 demonstrates that these relations give a pitch angle of arms of about 22° for our Galaxy, and this value is probably observed in its outer regions.

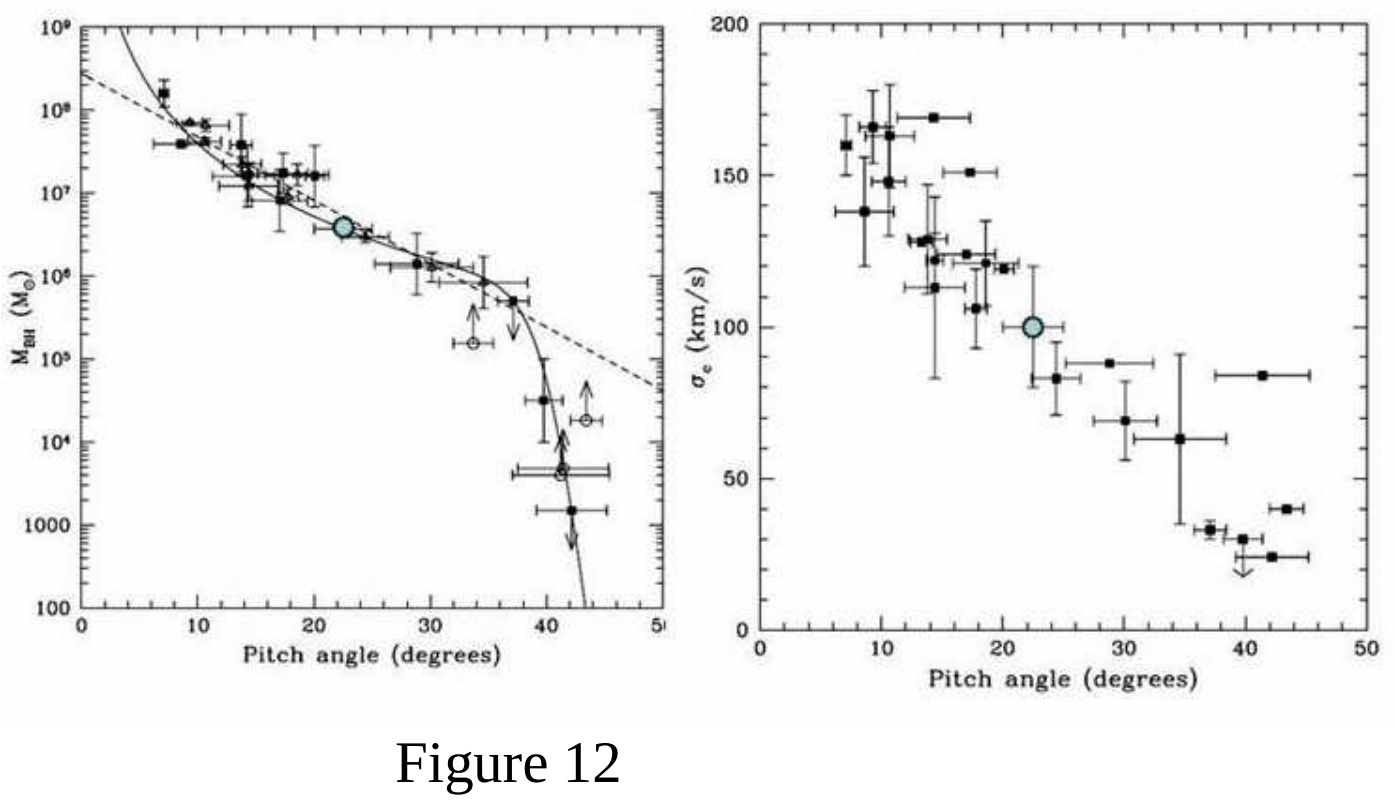

Figure 12

The relation between the spiral-arm pitch angle on the mass of the central black hole (left) and on the central velocity dispersion (right). The data were taken from [51], with the data for our Galaxy (the large circle) added.

Figure13 shows that the scheme of the Galaxy's outer arms from Levine et al. [25, 26] is compatible with the kneed structure of the spiral pattern of the Galaxy in its outer parts as well as with an increased pitch angle of the arms (to 20°--25°). These outer arms continue our four arms (it should be reminded that their pitch angle is 12°); the Cygnus arm, which deviates from the symmetry with the Carina arm in the I quadrant and obviously exhibits a bend between I and II quadrants, returns to its ``proper way'' in the IV quadrant and continues a regular logarithmic arm for a certain distance, but with a somewhat larger pitch angle. The Perseus arm also finds its continuation; it is apparent however that this arm as well as the Cygnus arm follow straights lines in their outmost parts.

This fact, along with the bends of both arms, the known gap in the Perseus arm towards the anticenter, and a straight spur from the Cygnus arm in the I and II quadrants (Fig. 13), can be considered indications of our Galaxy belonging to the M101-type multi-arm galaxies.

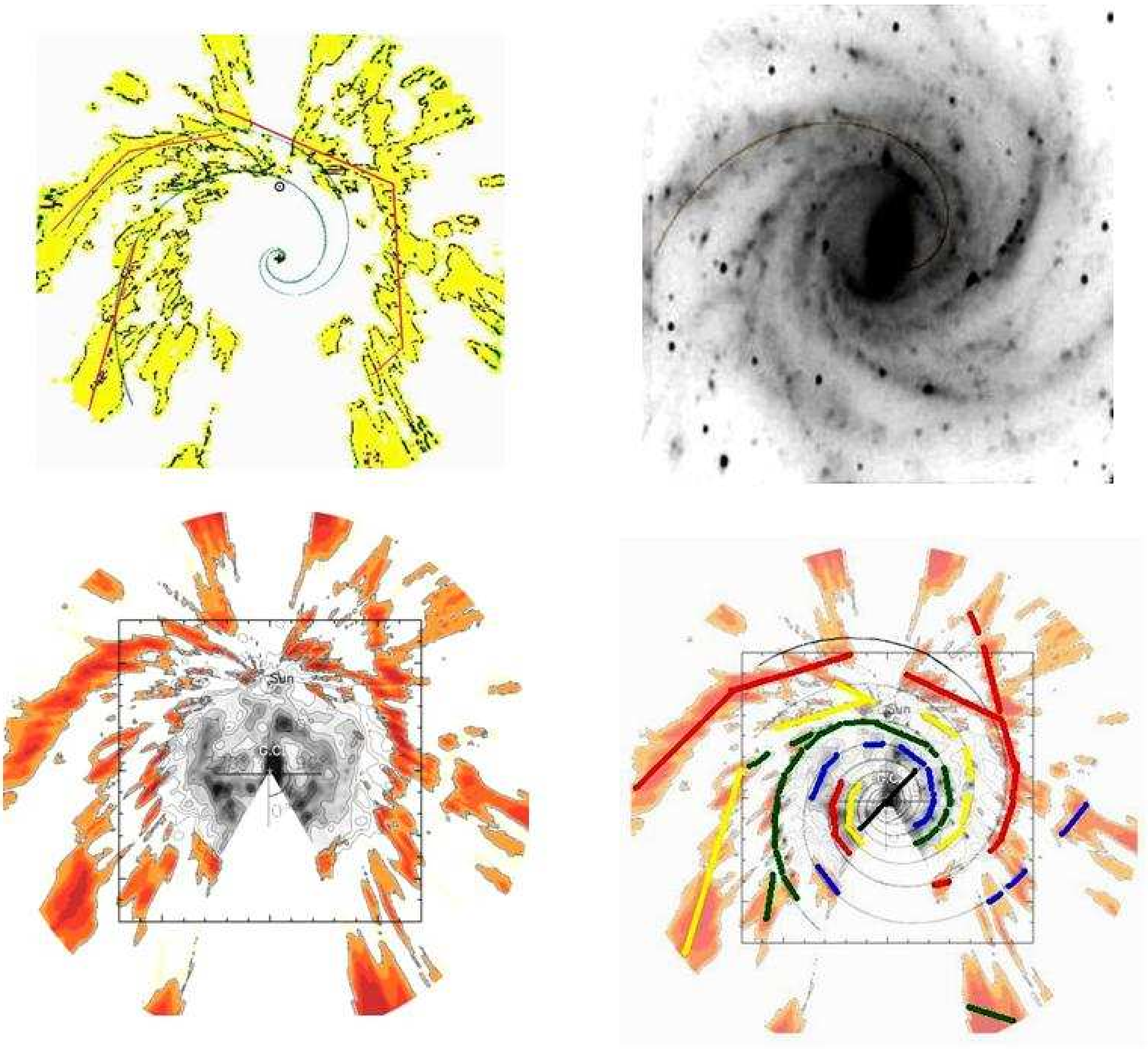

Figure 13

Top left: an attempt to plot kneed segments and arms with a large pitch angle (26º).
Top right: a fragment of a spiral with a pitch angle of 26° overlaid on one of the arms of the NGC 2336 galaxy.
Bottom left: the combined presentation of the neutral and molecular hydrogen distribution in the Galaxy (a combination of data from [26] and [31]).
Bottom right: the possible kneed structure of the Galaxy's spiral arms.

However, it is possible to notice from Fig. 13 that a continuation of the two arms approximately along logarithmic spirals but with a larger pitch angle is not excluded. This possibility is in agreement with the relation between the galaxy's rate of shear S and its pitch angle (*PA*) found by Seigar et al. [52]. The authors of the cited paper derived the following relation from rotation curves of galaxies:

$PA$ = 64-73 $S$, where the rate of shear $S$ is:
$S=A/\Omega =A/(A-B)$,
$A$ and $B$ being Oort's constants.

Assuming $A = 15$ *and* $B = -10$ *km/s/kpc* for our Galaxy, we find
$S = 0.60$, and $PA = 20°$.

The Oort's constants were determined for the solar vicinity (i.e. for galactocentric distances about 6--10~kpc), where the pitch angle is definitely smaller. However, for a flat rotation curve (observed, for our Galaxy, approximately starting at ~5 kpc from the center), we have $A = -B$ and thus $S = 0.5$, so that we get PA = 28°, using the formula from Seigar et al. [52]. A similar value of the pitch angle can be supposed for distances in excess of 10 –12 kpc using the data from Levine et al. [26] for the continuation of the Outer (Cygnus) arm in III quadrant (Fig. 13).

Thus, we may say that for our Galaxy the large pitch angle at large distances from the center is probably observed, in agreement with the relations from [51] and [52], which based on observations of other galaxies. This angle should increase when the rotation curve becomes flat, and such an increase can be assumed for several galaxies, like NGC 2336 (Fig. 13). However, the rotation curve of our Galaxy becomes flat approximately from the distance of about 5 kpc from the center, though the pitch angle of the arms in the inner Galaxy is surely no more than 12°.

## 8. CONCLUSIONS

It seems that the notion of regular logarithmic spirals in the inner Galaxy becoming kneed arms (consisting of straight fragments) in the outer Galaxy is most plausible. Our star system is a spiral galaxy resembling (among the most well-known systems) M101 but with a lower activity of star formation. By certain features, it also resembles M31 where the star formation is still less active, but the Milky Way differs from these two galaxy by the presence of a (relatively small) bar. Figure 14 displays images of the galaxies best resembling our star system, taken from the DSS and reduced to the face-on appearance. The classification of these galaxies is from the Sandage--Bedke atlas [53].

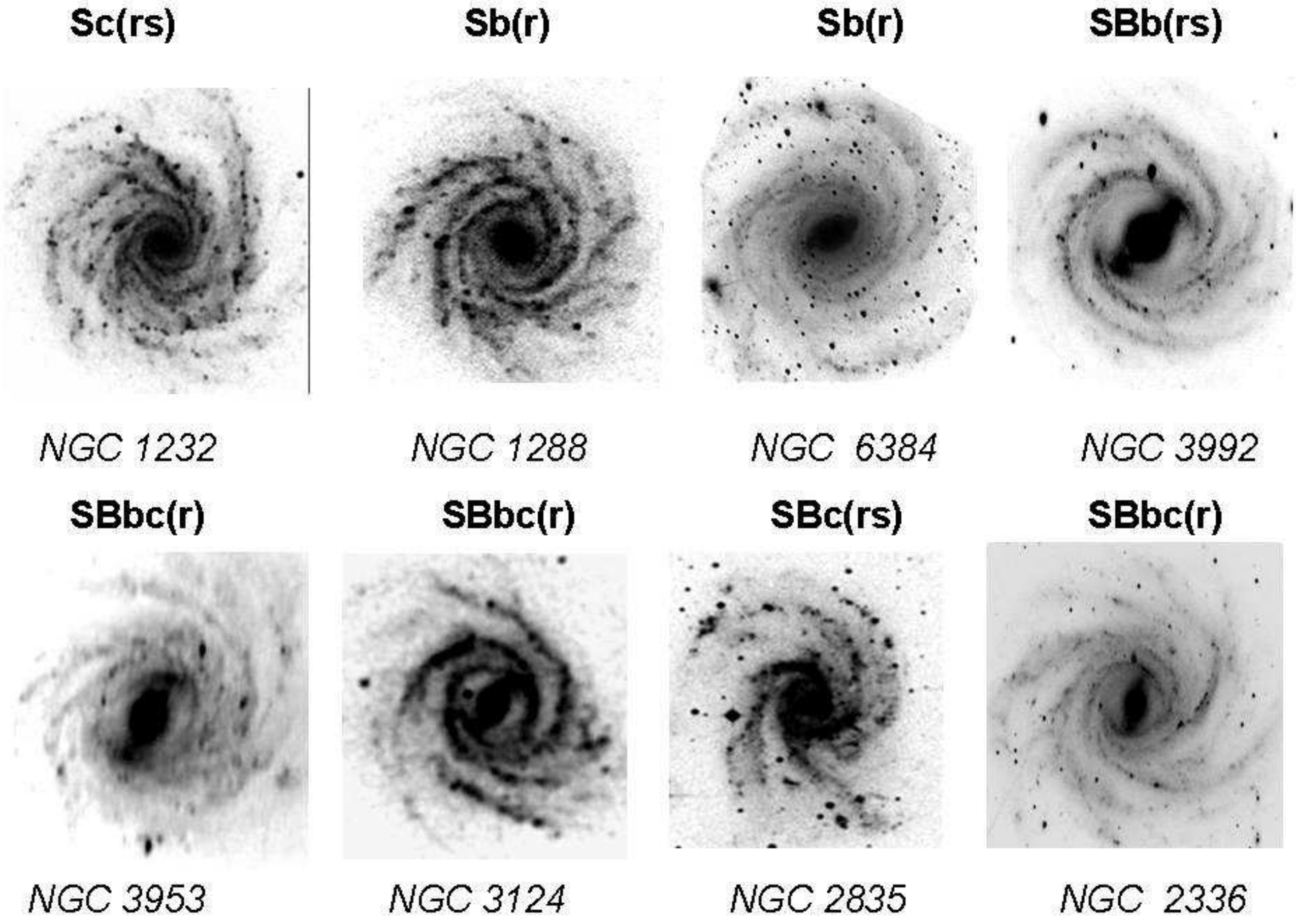

Figure 14

Galaxies resembling our Galaxy

Levine et al. [25, 26], who were the first to identify continuations of spiral arms at large distances from the center, came to a conclusion that our Galaxy was an asymmetric multi-arm spiral. The kneed shape of its outer arms and the straight shape of some fragments of its inner arms evidence for such arms being transient features formed due to gravitational instabilities in the gas disk. The high velocities of some star-formation regions recently found from VLBA data [29] are also difficult to explain in the frame of the theory of spiral arms as grand-design density waves.

However, the inner arms are more symmetric and obviously can be represented with logarithmic spirals. The presence of a bar and regular distances between the superclouds in the two arms which are best expressed in HI, evidence for these arms being related to spiral density waves. At distances to the center below the solar value, we have to choose between the four-arm scheme of the spiral arms with a pitch angle about 12° and a two-arm scheme with a pitch angle about 6°. The first option is much more probable; it is closer to the mean value for spiral arms of galaxies, while the angle of 6° is almost extremely small, such angles are observed only for galaxies with high concentration of mass to their centers that have a well-developed bulge (the *Sa* Hubble type), which is not the case for our Galaxy.

The distances between the HI superclouds in the two arms that are the brightest in neutral hydrogen, the Carina and Cygnus arms, concentrate to two preferred values, a possible indication that there is the regular magnetic field along the arm and the corresponding magneto-gravitational instability is responsible for formation of quasi-regular chains of gas superclouds and star complexes. It should be reminded that the equal distances between star complexes were found in M31 just in the arm [7, 8], where the presence of a regular magnetic field has been known for a long time [43].

The conclusion made by de Vaucouleurs [10] more than 40 years ago was that our Galaxy may resemble by its structure the galaxy NGC 4303 (M61), classified by him as SAB(rs). This conclusion was based on limited data on inner regions of the Milky Way. Note that M61 is a good example of a galaxy with kneed spiral arms. However, our Galaxy resembles stronger the galaxies like NGC 3124, NGC 3992 or NGC 2336, whose inner (regular and symmetric) arms originate near the tips of a small bar, but then become rather kneed, and the whole structure at large distances from the center is far from symmetry. Note however that far from the center of our Galaxy, we may discuss only the morphology of the arms traced with neutral hydrogen, which often extends to distances far beyond the optically visible structure of a spiral galaxy.

ACKNOWLEDGEMENTS

I am grateful to B.G.Elmegreen, A.S.Rastorguev, V.P.Arkhipova and A.D.Chernin for their helpful comments; to E.Yu.Efremov for preparing and improving some of the figures. This study was supported by the Russian Foundation for Basic Research through grant No.10--02--00178.